\documentclass[12pt,preprint]{emulateapj}
\usepackage{graphicx}

\shorttitle{Exoplanetary Transit Constraints}
\shortauthors{Stephen R. Kane \& Kaspar von Braun}
\slugcomment{Accepted for publication in PASP}

\begin{document}

\title{Exoplanetary Transit Constraints Based Upon Secondary Eclipse
  Observations}
\author{Stephen R. Kane}
\author{Kaspar von Braun}
\affil{NASA Exoplanet Science Institute, Caltech, MS 100-22, 770 South
  Wilson Avenue, Pasadena, CA 91125, USA}
\email{skane, kaspar@ipac.caltech.edu}


\begin{abstract}

Transiting extrasolar planets provide an opportunity to study the
mass-radius relation of planets as well as their internal
structure. The existence of a secondary eclipse enables further study
of the thermal properties of the the planet by observing at infrared
wavelengths. The probability of an observable secondary eclipse
depends upon the orbital parameters of the planet, particularly
eccentricity and argument of periastron. Here we provide analytical
expressions for these probabilities, investigate their properties, and
calculate their values for the known extrasolar planets. We
furthermore quantitatively discuss constraints on existence and
observability of primary transits if a secondary eclipse is
observed. Finally, we calculate the a-posteriori transit probabilities
of the known extrasolar planets, and we present several case studies
in which orbital constraints resulting from the presence of a
secondary eclipse may be applied in observing campaigns.

\end{abstract}



\section{Introduction}
\label{intro}

Transiting planet discoveries have become an integral component of the
extrasolar planets field, with discoveries taking place at an ever
increasing rate. The additional information provided by the detection
of a transit to the overall understanding of both the orbital elements
and the planetary properties is invaluable. In particular, data
acquired through the observations of secondary eclipses has allowed an
unprecedented insight into the analysis of planetary atmospheres
\citep{bur05,bur06,gri08,knu07,wil06}. These secondary eclipses have
been observed most notably from space \citep{dem06,dem07a,dem07b} but
have also been detected from the ground \citep{dem09,gil09,sin09}.

The effect of orbital parameters upon the probability of an observable
primary transit has been discussed in detail by such papers as
\citet{bar07b} and \citet{bur08}. The effects of the eccentricity and
argument of periastron have a considerable effect on this probability
for many of the known planets discovered through the radial velocity
method \citep{kan08}. An example of this is the planet orbiting HD
17156 whose relatively large eccentricity results in a high primary
transit probability. Subsequent observing campaigns confirmed that
this planet does indeed transit its parent star \citep{bar07a,win09}.

For a given eccentricity, the inverse case of a high primary transit
probability is that of a high secondary eclipse probability. The case
of HD~80606b \citep{nae01} is a spectacular example with an orbital
eccentricity of $\sim 0.93$ and argument of periastron of $\sim
300\degr$, resulting in an especially high secondary eclipse
probability, despite an orbital period of more than 111 days. The
secondary eclipse was successfully detected by \citet{lau09} using the
Spitzer Space Telescope's Infrared Array Camera. With the observation
of a seconday eclipse, efforts were undertaken to determine if the
planet also produces an observable primary transit. These efforts were
eventually fruitful with the confirmation of a primary transit
\citep{fos09,gar09,mou09}.

Despite a favourable orbital inclination, merely the presence of
either a secondary eclipse or a primary transit does not necessarily
imply that its counterpart will be detectable. If a secondary eclipse
is observed for a known radial velocity planet (as was the case for
HD~80606b) then the likelihood of a primary transit is
increased. However, exactly how likely this primary transit is will
inform the justification for mounting an exhaustive follow-up
campaign. Here we calculate secondary eclipse probablities for known
radial velocity planets and show how improved estimates on primary
transit probabilities and predicted transit mid-points can be placed
if a secondary eclipse is observed. We further demonstrate this with
several case studies which apply these constraints to some of the
known radial velocity planets.


\section{Secondary Eclipses}

In this section we calculate secondary eclipse probablities and apply
it to the known radial velocity planets. We also discuss the
limitations of impact parameter measurements and subsequent
uncertainties in the estimation of the orbital inclination from
secondary eclipse observations. In this and all subsequent sections,
we use $t$ and $e$ as subscripts for (primary) transit and (secondary)
eclipse, respectively.


\subsection{Geometric Eclipse Probability}

\begin{figure*}
  \includegraphics[width=17.3cm]{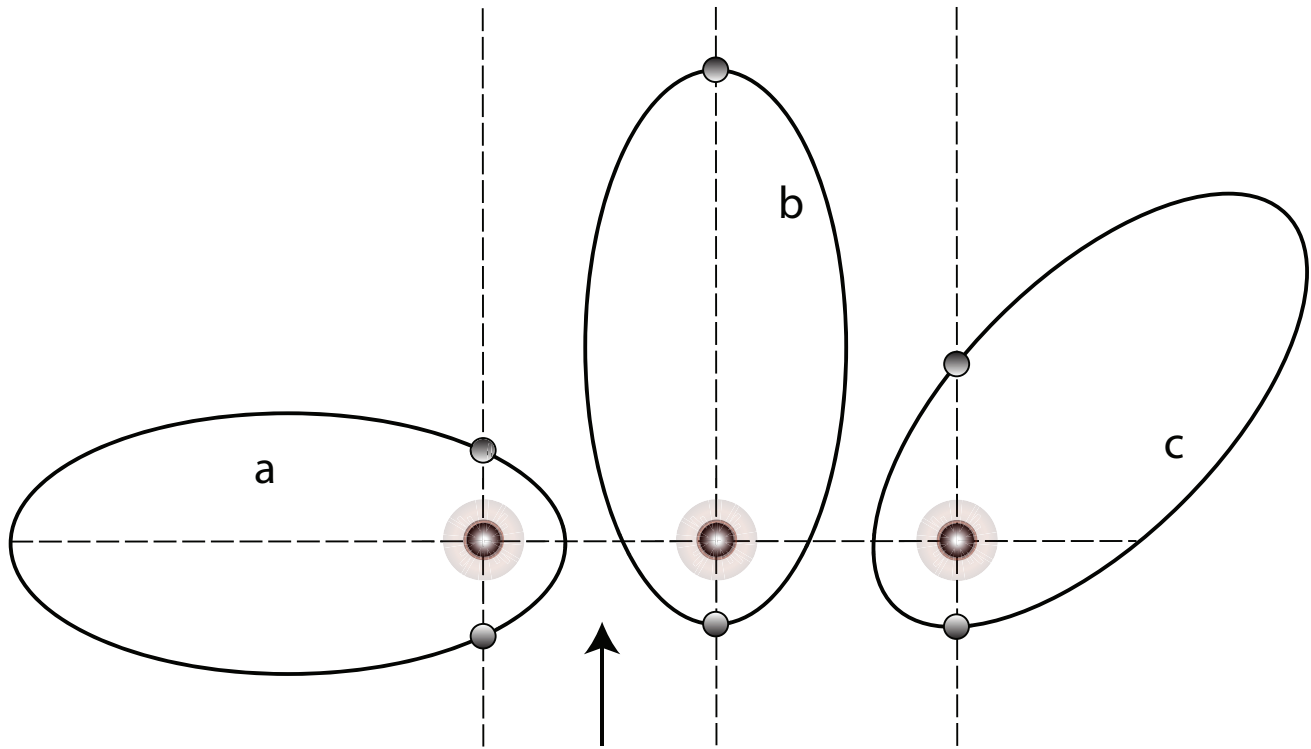}
  \caption{Top-down view of three different orbital configations of an
    eccentric orbit, with the arrow indicating the line of sight of an
    observer. The periastron arguments of orbits a, b, and c are
    $\pi$, $\pi / 2$, and $\pi / 4$ respectively. The star--planet
    distance both in front of and behind the star is highly dependent
    upon this periastron argument.}
  \label{orbits}
\end{figure*}

The geometric transit probability, $P_t$, is often approximated as the
ratio of the radius of the parent star, $R_\star$, to the semi-major
axis of the planetary orbit, $a$. A more thorough consideration of the
orbital parameters shows that the probability of a primary transit can
be more accurately described by
\begin{equation}
  P_t = \frac{(R_p + R_\star)(1 + e \cos (\pi/2 - \omega))}{a (1 - e^2)}
  \label{tranprob}
\end{equation}
where $R_p$ is the radius of the planet, $e$ is the eccentricity, and
$\omega$ is the argument of periastron. The effects of $e$ and
$\omega$ upon the transit probability can be considerable, as shown by
\citet{kan08}. This probability peaks where $\omega = \pi / 2$.

The true anomaly, $f$, is defined as the angle between the current
position of the planet in its orbit and the direction of
periapsis. The location in the orbit at which the planet crosses a
plane between the host star and the observer which is perpendicular to
the orbit is where a primary transit is possible to occur, and is
where $\omega + f = \pi / 2$ \citep{kan07}. If we extend the
star--observer plane beyond the host star, the location where the
planet crosses the plane on the far side of the star is where $\omega
+ f = 3 \pi / 2$ and is the location where it is possible for for a
secondary eclipse to occur. Hence the geometric eclipse probability,
$P_e$, is given by
\begin{equation}
  P_e = \frac{(R_p + R_\star)(1 + e \cos (3\pi/2 - \omega))}{a (1 -
    e^2)}.
  \label{eclprob}
\end{equation}
This probability peaks where $\omega = 3 \pi / 2$.

Depending upon the orientation of an eccentric orbit, it is possible
for planets to have both a higher transit and eclipse probability than
that produced by a circular orbit with the same period. In other
words, there is a range of periastron arguments for which the
projected star--planet separation:
\begin{equation}
  r = \frac{a (1 - e^2)}{1 + e \cos f}
  \label{separation}
\end{equation}
is smaller than that for a circular orbit with the same period both in
front of and behind the star along the line of sight. This is
demonstrated in Figure \ref{orbits}, which shows three different
orbital orientations relative to the line of sight of an observer
(indicated by the arrow). The periastron arguments of orbits a, b, and
c are $\pi$, $\pi / 2$, and $\pi / 4$ respectively. This visualization
of the various configurations clearly shows how the star--planet
separation along the line of sight is dependent on the argument of
periastron and subsequently affects both the transit and eclipse
probabilities.

\begin{figure}
  \includegraphics[angle=270,width=8.2cm]{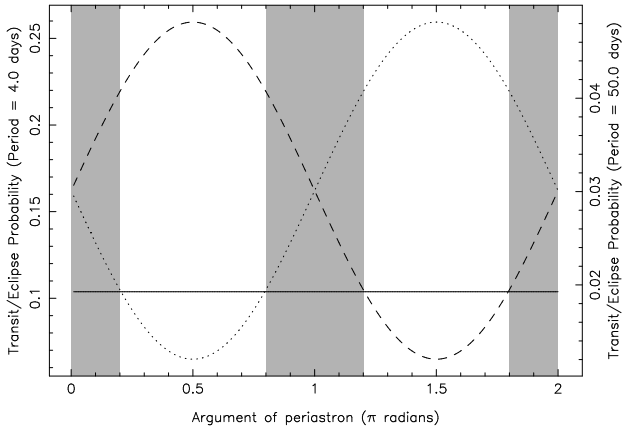}
  \caption{Dependence of geometric transit (dashed line) and eclipse
    (dotted line) probability on the argument of periastron for an
    eccentricity of 0.6 (see Equations \ref{tranprob} and
    \ref{eclprob}). The solid line indicates the probabilities for a
    circular orbit with the same orbital period. These are plotted for
    periods of 4.0 days (left ordinate) and 50.0 days (right
    ordinate). Stellar and planetary radii are assumed to be a Jupiter
    and solar radius, respectively. The shaded regions represent those
    ranges of periastron arguments for which both the transit and
    eclipse probabilities exceed that of a circular orbit.}
  \label{regions}
\end{figure}

Figure \ref{regions} shows the dependence of transit and eclipse
probability upon periastron argument for an eccentricity of 0.6, along
with those regions for which both probabilities exceed that of a
corresponding circular orbit (shaded regions). The stellar mass and
radius are assumed to be solar mass and radius respectively, whilst
the planet is assumed to have a Jupiter radius. For this particular
eccentricity, the shaded regions account for 40\% of the possible
periastron arguments. It is often non-intuitive that there exist
orbital configurations where both probabilities are enhanced since
they are often assumed to have inverse relation to each other.

\begin{figure}
  \includegraphics[angle=270,width=8.2cm]{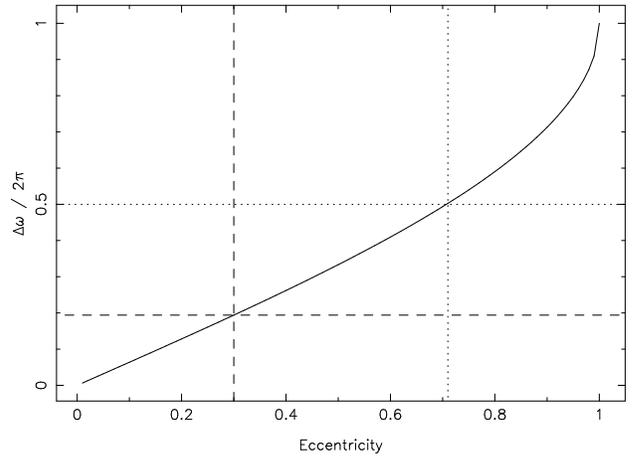}
  \caption{Dependence of the fractional range of periastron arguments
    for which both the transit and eclipse probabilities exceed that
    of a circular orbit with the same period (shaded regions in Figure
    \ref{regions}) on the orbital eccentricity. The dashed lines
    correspond to $e = 0.3$ and the dotted lines correspond to $\Delta
    \omega / 2 \pi = 0.5$.}
  \label{deltaomega}
\end{figure}

The total size of the shaded regions depends upon the eccentricity of
the orbit and can be calculated analytically by considered those
values of $\omega$ for which the transit probability of a circular
orbit equals that of an eccentric orbit for the same period. Using
Equation \ref{separation}, this is where
\begin{equation}
  \frac{a (1 - e_0^2)}{1 + e_0 \cos f} = \frac{a (1 - e^2)}{1 + e \cos
    f}
\end{equation}
which yields
\begin{equation}
  f = \pm \cos^{-1} (-e)
\end{equation}
when $e_0 = 0$ (circular orbit). The values of $\omega$ which define
the boundaries of the shaded regions in Figure \ref{regions} are
then given by $\omega + f = \pi / 2$ and $\omega + f = 3 \pi / 2$ for
primary transits and secondary eclipses respectively. The total size
of the shaded regions $\Delta \omega$ can then be expressed as
\begin{equation}
  \Delta \omega = 4 \cos^{-1} (-e) - 2 \pi.
\end{equation}
Figure \ref{deltaomega} shows the dependence of the value of $\Delta
\omega$ on the eccentricity of the orbit. According to \citet{bur08},
the mean eccentricity of the known extra-solar planets is $\sim 0.3$
for orbital periods greater than 10 days. This is indicated on Figure
\ref{deltaomega} as a dashed line which shows that almost 20\% of the
$\omega$ values for this eccentricity will have higher transit/eclipse
probabilities than for a circular orbit with the same period. The
eccentricity for which half of the $\omega$ values fall into this
category is also shown as a dotted line. This occurs when the
magnitude of the eccentricity exceeds $\sim 0.71$. This can be used to
identify planets for which observations at both predicted transit and
eclipse times would be beneficial to make efficient use of observing
time. In general, if the value of $e$ has been measured, then the
value of $\omega$ will also have an associated estimate which will
define exactly where on Figures \ref{regions} and \ref{deltaomega} the
planet lies.  It should also be noted that a high probability of both
transit and eclipse provides access to greater science
opportunities. A transit passage primarily enables measurement of the
planetary radius, whereas an eclipse passage primarily enables
measurement of the planetary flux which, when combined with planetary
atmosphere models, can lead to an estimate of the planetary
temperature.

\begin{figure}
  \includegraphics[angle=270,width=8.2cm]{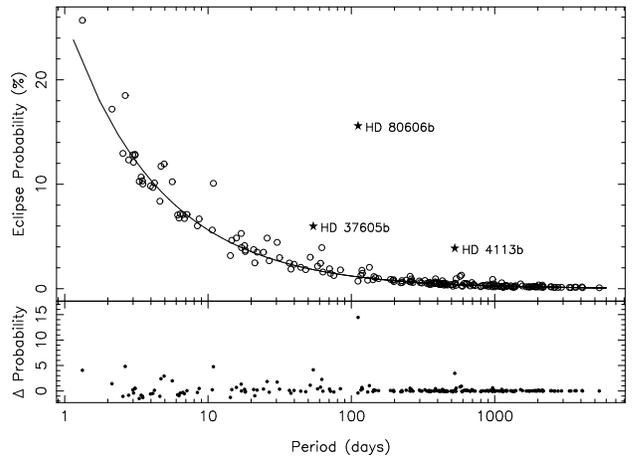}
  \caption{The geometric eclipse probability for a circular orbit with
    the published period (solid curve) along with the eclipse
    probability for 203 RV planets from \citet{but06} calculated from
    their orbital parameters (open circles). HD~80606b, HD~4113b, and
    HD~37605b are indicated by 5-pointed stars as examples of
    particularly high eclipse probabilities. The lower panel plots the
    difference in $P_e$ between the actual orbit and a hypothetical
    circular orbit with the same period for each of the planets.}
  \label{gep}
\end{figure}

\citet{kan08} used the orbital parameters provided by \citet{but06} to
calculate the primary transit probabilities for 203 planets. These
calculations took into account the eccentricity and argument of
periastron to demonstrate the inflated probabilities that can occur as
a result. In Figure \ref{gep} we show the results of performing a
similar calculation using the same data to estimate secondary eclipse
probabilities. We assume a Jupiter and Solar radius for the values of
$R_p$ and $R_\star$ respectively to provide ease of comparison with
the eclipse probability of a circular orbit with the same period,
shown as a solid curve. Note that there are a handful of M dwarfs in
the sample whose eclipse probabilities will be a factor of two lower
than those shown in the figure dur to the assumption regarding stellar
radius.

The secondary eclipse probability of HD~80606b is labelled in Figure
\ref{gep}. Since HD~80606 is of solar-type (G5V) and the planet is
approximately the same size as Jupiter then the indicated probability
is close to the true probability for this planet. The residuals show
that the increase in eclipse probability for this planet compared to a
circular orbit with the same period is $> 15$\%. Apart from HD~80606b
and a few outliers at periods $< 100$ days, another example of an
unexpectedly high $P_t$ in this plot is the long-period planet
HD~4113b \citep{tam08}.  This planet has a period of 526.62 days and
an eccentricity of 0.903. Such a high eccentricity means that,
according to Figure \ref{deltaomega}, $\sim 70$\% of the possible
values of $\omega$ will result in both a higher transit and eclipse
probability than if the eccentricity were zero. Thus, even though the
periastron argument is $\omega = 317.7\degr$, the secondary eclipse
probability is still raised by 3.5\% compared to the corresponding
circular orbit.


\subsection{Impact Parameter and Inclination}
\label{impactpar}

Here we show how the existence of a secondary eclipse can constrain
values of the impact parameter and orbital inclination angle.  The
dimensionless impact parameter, $b$, of an exoplanetary transit is
defined as the projected separation of the planet and star centers at
the point of mid-transit. Thus $b$ is related to the inclination, $i$,
by
\begin{equation}
  b \equiv \frac{r}{R_\star} \cos i
  \label{impact}
\end{equation}
such that $0 \le b \le 1$. In Equation \ref{impact}, we have
generalised from a circular to an eccentric orbit by replacing $a$
with $r$. As discussed in detail by \citet{sea03}, the measured impact
parameter is highly dependant upon the shape of the light curve, in
particular the duration of ingress and egress. \citet{sea03} also
point out that $b$ will typically be underestimated when there is
relatively low signal-to-noise in the transit data. This will be true
for almost all ground-based observations of secondary eclipses, and is
even sometimes true for Spitzer secondary eclipse observations.

If we assume that the maximum secondary eclipse depth is achieved (the
full disc of the planet passes behind the star), then $b$ can be as
high as $(R_\star - R_p) / R_\star$. The orbital inclination will then
be as low as
\begin{equation}
  i = \cos^{-1} \left( \frac{R_\star - R_p}{r_e} \right)
  \label{inclination}
\end{equation}
where $r_e$ is the star--planet separation at eclipse (see Equation
\ref{separation}). For a Jupiter-type planet orbiting a solar-type
star with a period of 4 days, this results in a maximum impact
parameter of $b \sim 0.9$ and hence a minimum orbital inclination of
$i \sim 85\degr$. For HD~80606b, the orbital parameters of $e =
0.9336$ and $\omega = 300.4977\degr$ \citep{fos09}, lead to a
star--planet separation of $r_e = 0.032$~AU at mid-eclipse and $r_t =
0.297$~AU at mid-transit. Adopting the stellar and planetary radii
used by \citet{fos09}, the minimum inclination for this planet is $i =
82.5\degr$.


\section{Primary Transit Constraints}

In this section we show the impact the detection of a secondary
eclipse has on the predictions regarding an observable primary transit
of the planet.


\subsection{Geometric Transit Probability}
\label{gtp}

In most cases, an estimate of a planet's primary transit probability
is made with no knowledge of the planet's orbital
inclination. However, the constraints on the inclination discussed in
the previous section allow for an improved estimate of the primary
transit probability. Since the minimum inclination can be calculated
from the presence of a secondary eclipse, the probability of a primary
transit is given by
\begin{equation}
  P_t \geq \frac{R_p + R_\star}{r_t \cos i}.
\end{equation}
Note the inequality used in this equation to make it clear that this
represents a lower limit on the true transit probability. Using
Equation \ref{inclination}, this probability may be re-expressed as
\begin{equation}
  P_t \geq \frac{(R_\star + R_p)}{(R_\star - R_p)} \frac{r_e}{r_t}
\end{equation}
thus removing the inclination dependence. Generally speaking, if $r_e
> r_t$ then a secondary eclipse detection almost guarantees that a
primary transit will also be observable. By substituting Equation
\ref{separation} and using trigonometric identities, the probability
may be re-expressed once again as
\begin{equation}
  P_t \geq \frac{(R_\star + R_p)}{(R_\star - R_p)} \frac{(1 + e \sin
    \omega)}{(1 - e \sin \omega)}
  \label{newprob}
\end{equation}
which removes dependence upon semi-major axis or period, and replaces
these with the dependence upon eccentricity and periastron argument.

\begin{figure*}
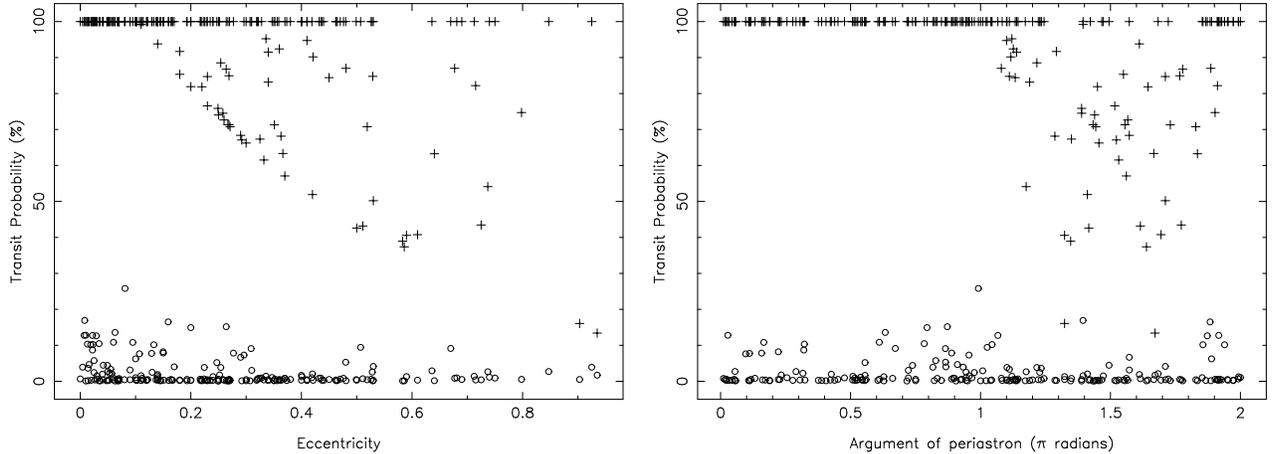

  \begin{center}
    \begin{tabular}{cc}
      \includegraphics[angle=270,width=8.2cm]{f5a.eps} &
      \includegraphics[angle=270,width=8.2cm]{f5b.eps} \\
    \end{tabular}
  \end{center}
  \caption{The original (Equation \ref{tranprob}) and revised
    (Equation \ref{newprob}) primary transit probabilities for 203
    planets from the \citet{but06} catalog, plotted as a function of
    eccentricity and argument of periastron. The original
    probabilities are shown as open circles and the revised (when a
    secondary eclipse is detected) probabilities are shown as
    crosses.}
  \label{newgtp}
\end{figure*}

Figure \ref{newgtp} demonstrates the improvement one gains in the
primary transit probability if a secondary eclipse is detected as a
function of both eccentricity and periastron argument. This
demonstration uses the same orbital parameters for the 203 planets
shown in Figure \ref{gep} as provided by \citet{but06}. As for Figure
\ref{gep}, Jupiter and solar radii are assumed for the planetary and
host star radii respectively. The open circles are the transit
probabilities calculated using Equation \ref{tranprob}, which assumes
no prior knowledge regarding the inclination of the planetary
orbit. The crosses are the revised transit probabilities calculated
from Equation \ref{newprob}, which assumes that a secondary eclipse
has been observed. The 50 highest eclipse probability planets from
this sample are tabulated in Table \ref{top50} along with basic
orbital parameters and their original and revised transit
probabilities.

Figure \ref{newgtp} and Table \ref{top50} show that for many cases the
transit probability is elevated to 100\% when a secondary eclipse is
observed, indeed this is true for almost 75\% of the planets included
in this sample. The left panel indicates that the transit probability
for the stars with a postulated observation of secondary eclipse
(crosses) fans out for $e>0.1$ with an envelope whose lower edge
appears linear with a negative slope. The right panel demonstrates
that if the argument of periastron is close to $\pi / 2$, then one can
be assured of an observable primary transit based upon a secondary
eclipse. However, if the argument of periastron is close to $3 \pi /
2$ then a non-circular orbit will correspondingly reduce this
improvement to the transit probability. This region corresponds to the
region of high secondary eclipse probability shown in Figure
\ref{regions}.

\begin{table}
  \begin{center}
    \caption{The 50 highest eclipse probability planets from the
      \citet{but06} sample. Tabulated are the period, $P$,
      eccentricity, $e$, and periastron argument, $\omega$, as well as
      the transit probability, $P_t$, eclipse probability, $P_e$, and
      the revised transit probabilities, $P_t'$, calculated from
      Equation \ref{newprob}.}
    \label{top50}
    \begin{tabular}{@{}lcccccc}
      \hline
      Planet & $P$ (d) & $e$ & $\omega$ ($\degr$) & $P_t$
      (\%) & $P_e$ (\%) & $P_t'$ (\%) \\
      \hline
HD 41004 B b  &      1.33 &  0.08 &  178.50 &   25.81 &   25.70 &  100.00 \\
GJ 436 b      &      2.64 &  0.16 &  339.00 &   16.50 &   18.50 &  100.00 \\
HD 86081 b    &      2.14 &  0.01 &  251.00 &   16.93 &   17.19 &  100.00 \\
HD 80606 b    &    111.45 &  0.93 &  300.89 &    1.71 &   15.58 &   13.41 \\
HD 73256 b    &      2.55 &  0.03 &  337.00 &   12.66 &   12.95 &  100.00 \\
HD 179949 b   &      3.09 &  0.02 &  192.00 &   12.74 &   12.86 &  100.00 \\
HD 83443 b    &      2.99 &  0.01 &  345.00 &   12.77 &   12.82 &  100.00 \\
HD 187123 b   &      3.10 &  0.01 &    5.03 &   12.81 &   12.78 &  100.00 \\
55 Cnc e      &      2.80 &  0.26 &  157.00 &   15.17 &   12.33 &  100.00 \\
HD 46375 b    &      3.02 &  0.06 &  114.00 &   13.58 &   12.11 &  100.00 \\
HD 49674 b    &      4.94 &  0.29 &  283.00 &    6.68 &   11.94 &   68.37 \\
GJ 674 b      &      4.69 &  0.20 &  143.00 &   14.93 &   11.72 &  100.00 \\
HD 88133 b    &      3.42 &  0.13 &  349.00 &   10.16 &   10.68 &  100.00 \\
BD-10 3166 b  &      3.49 &  0.02 &  334.00 &   10.15 &   10.32 &  100.00 \\
tau Boo b     &      3.31 &  0.02 &  188.00 &   10.21 &   10.27 &  100.00 \\
HAT-P-2 b     &      5.63 &  0.51 &  184.60 &    9.44 &   10.24 &  100.00 \\
51 Peg b      &      4.23 &  0.01 &   58.00 &   10.35 &   10.12 &  100.00 \\
HD 108147 b   &     10.90 &  0.53 &  308.00 &    4.14 &   10.09 &   50.21 \\
HD 75289 b    &      3.51 &  0.03 &  141.00 &   10.47 &   10.03 &  100.00 \\
HD 76700 b    &      3.97 &  0.09 &   29.90 &   10.82 &    9.84 &  100.00 \\
HD 102195 b   &      4.12 &  0.06 &  109.90 &   10.85 &    9.69 &  100.00 \\
upsilon And d &      4.62 &  0.02 &   57.60 &    8.69 &    8.38 &  100.00 \\
HD 168746 b   &      6.40 &  0.11 &   17.40 &    7.63 &    7.16 &  100.00 \\
HD 217107 b   &      7.13 &  0.13 &   20.00 &    7.76 &    7.11 &  100.00 \\
HIP 14810 b   &      6.67 &  0.15 &  160.00 &    7.86 &    7.10 &  100.00 \\
HD 118203 b   &      6.13 &  0.31 &  155.70 &    9.11 &    7.05 &  100.00 \\
HD 68988 b    &      6.28 &  0.15 &   40.00 &    8.20 &    6.76 &  100.00 \\
HD 185269 b   &      6.84 &  0.30 &  172.00 &    7.30 &    6.72 &  100.00 \\
HD 69830 b    &      8.67 &  0.10 &  340.00 &    6.24 &    6.68 &  100.00 \\
HD 162020 b   &      8.43 &  0.28 &   28.40 &    7.84 &    6.02 &  100.00 \\
HD 37605 b    &     54.23 &  0.74 &  211.60 &    2.64 &    5.97 &   54.12 \\
HD 130322 b   &     10.71 &  0.03 &  149.00 &    5.76 &    5.62 &  100.00 \\
HD 99492 b    &     17.04 &  0.25 &  219.00 &    3.83 &    5.29 &   88.53 \\
HD 13445 b    &     15.76 &  0.04 &  269.00 &    4.47 &    4.85 &  100.00 \\
HD 117618 b   &     25.83 &  0.42 &  254.00 &    2.06 &    4.85 &   51.92 \\
55 Cnc b      &     14.65 &  0.02 &  164.00 &    4.67 &    4.63 &  100.00 \\
GJ 876 c      &     30.34 &  0.22 &  198.30 &    3.85 &    4.44 &  100.00 \\
HD 27894 b    &     17.99 &  0.05 &  132.90 &    4.43 &    4.12 &  100.00 \\
HD 3651 b     &     62.24 &  0.59 &  238.20 &    1.30 &    3.93 &   40.58 \\
HD 190360 c   &     17.11 &  0.00 &  168.00 &    3.94 &    3.93 &  100.00 \\
HD 4113 b     &    526.62 &  0.90 &  238.20 &    0.51 &    3.86 &   16.08 \\
HD 102117 b   &     20.81 &  0.09 &  283.00 &    3.14 &    3.74 &  100.00 \\
HD 195019 b   &     18.20 &  0.01 &  222.00 &    3.62 &    3.69 &  100.00 \\
HD 33283 b    &     18.18 &  0.48 &  155.80 &    5.31 &    3.56 &  100.00 \\
HD 6434 b     &     22.00 &  0.17 &  156.00 &    4.02 &    3.50 &  100.00 \\
HD 192263 b   &     24.36 &  0.05 &  200.00 &    3.36 &    3.49 &  100.00 \\
HD 38529 b    &     14.31 &  0.25 &  100.00 &    5.21 &    3.17 &  100.00 \\
HD 74156 b    &     51.64 &  0.64 &  181.50 &    2.91 &    3.01 &  100.00 \\
HD 69830 c    &     31.56 &  0.13 &  221.00 &    2.51 &    2.97 &  100.00 \\
HD 224693 b   &     26.73 &  0.05 &   10.00 &    2.72 &    2.68 &  100.00 \\
      \hline
    \end{tabular}
  \end{center}
\end{table}


\subsection{Expected Primary Transit Time}
\label{trantime}

If a secondary eclipse is observed, then how long will it be before
the expected time of primary transit occurs? Can the necessary
observing resources be acquired or alerted in time to perform the
observations?  What will be the size of the transit window and can the
observations be justified with regards to the uncertainty in the
transit mid-point, the predicted transit duration, and the transit
probability? These are all fair questions to ask when planning a
follow-up campaign in the wake of a secondary eclipse detection,
especially in light of the deteriorating precision of the transit
ephemerides with time.

The predicted time of mid-transit can be calculated by utilising
Kepler's equations. Firstly, the eccentric anomaly is calculated from
the following relation to the true anomaly
\begin{equation}
  E = 2 \tan^{-1} \left( \sqrt{\frac{1-e}{1+e}} \tan \frac{f}{2}
  \right).
  \label{eccanom}
\end{equation}
The mean anomaly, $M$, which defines the time since last periapsis in
units of radians, is then computed by
\begin{equation}
  M = E - e \sin E
  \label{meananom}
\end{equation}
which can be converted to regular time units using
\begin{equation}
  t_M = \frac{P M}{2 \pi}.
  \label{meantime}
\end{equation}
By substituting $\omega + f = \pi / 2$ and $\omega + f = 3 \pi / 2$
into Equation \ref{eccanom}, we can calculate the predicted times of
primary transit, $t_t$, and secondary eclipse, $t_e$, respectively
\begin{eqnarray}
  t_t = t_{\mathrm{peri}} + t_M|_{(\omega + f = \pi / 2)}\\
  t_e = t_{\mathrm{peri}} + t_M|_{(\omega + f = 3 \pi / 2)}
\end{eqnarray}
where $t_{\mathrm{peri}}$ is the time of periastron passage. These
times can then be combined to yield the predicted time of primary
transit as a function of secondary eclipse
\begin{equation}
  t_t = t_e + \frac{P}{2 \pi} (M_t - M_e) + n P
  \label{timediff}
\end{equation}
where $M_t$ and $M_e$ are the mean anomalies for primary transit and
secondary eclipse respectively. This equation is true for $-\pi/2 \leq
\omega \leq \pi/2$, with an additional period needing to be added for
$\pi/2 \leq \omega \leq 3\pi/2$.  The term of $n \times P$ can be used
to calculate an ephemeris where $n$ is the number of complete orbits
one would like to consider. As expected, Equation \ref{timediff}
reduces to $t_t = t_e + P / 2$ when $e = 0$.

The uncertainties in the orbital parameters can be propogated through
these equations to determine the size of the transit window. Under
normal circumstances, the size of a transit window is most dependent
upon the uncertainty in the period and the time elapsed since last
observations were acquired. However, if a secondary eclipse has been
observed then the constraints on the window tighten and become
dominated by the eccentricity.


\section{Case Studies}

There are several specific cases of known exoplanets for which it is
useful to apply the principles described above. We consider some of
these cases here.


\subsection{HD 80606b}

HD~80606b \citep{nae01} is the first planet whose secondary eclipse
was discovered before its primary transit. As described in Section
\ref{impactpar}, if we used the revised orbital parameters by
\citet{fos09}, then the observation of a secondary eclipse places a
constraint of $i = 82.5\degr$ on the inclination of the orbit. The
equations in Section \ref{newgtp} yield an a-posteriori primary
transit probability of 13.4\%. For comparison, the probability of a
primary transit without any knowledge of the orbital inclination is
1.7\%.

Using the equations in Section \ref{trantime}, one can calculate the
time between secondary eclipse and primary transit to be only 5.86
days; small compared to the 111.43 day period. Additionally, the time
difference between secondary eclipse mid-point and periapsis is 0.12
days ($\sim 3$ hours); emphasizing the suitability of this system for
secondary eclipse detection. This time difference yields a predicted
transit mid-point of HJD 2454876.32 for the 5th primary transit after
the eclipse observed by \citet{lau09}, comparable to the observed
mid-transit time of HJD 2454876.344 by \citet{fos09}.


\subsection{HD 4113b}

The planet HD~4113b \citet{tam08} is a relatively long period (526.62
days) planet in an eccentric orbit ($e = 0.903$). The periastron
argument is $\omega = 317.7\degr$, which results in a secondary
eclipse probability of 3.9\%. In contrast, the transit probability for
this planet is only 0.5\%. If this planet were observed to undergo a
secondary eclipse, then the subsequent constraints upon the orbital
inclination increase the transit probability to an attractive 16.1\%.

The long period of this planet's orbit implies an expected transit
duration of $\sim 19$ hours, which produces the observational
challenge of attempting to observe either ingress or egress. If a
secondary eclipse is observed, then the next predicted transit
mid-point will occur only 19.55 days later. Although it is certainly
possible to marshall the needed observing resources within that
time-frame, a missed opportunity will require waiting an entire
complete period of 526.62 days before the next chance
arrives. Additionally, the similar orbital orientation of HD~4113b to
HD~80606b with respect to the observer means that the time between
eclipse and periapsis is only 1.72 days.


\subsection{HD 37605b}

The eccentric planet HD~37605b was discovered by \citet{coc04}. Since
then, the orbital parameters have been revised and published in the
catalog by \citet{but06}. The orbital period of 54.23 days,
eccentricity of 0.731, and periastron argument of $211.6\degr$ yield a
secondary eclipse probability of 6.0\%. A priori, the transit
probability of this planet is 2.6\%. The constraints placed upon the
inclination if a secondary eclipse is detected raises this probability
to an impressive 54.1\%.

The relatively short period of this planet make this an attractive
target for follow-up campaigns. The time period between periastron
passage and the secondary eclipse is 1.09 days. After the secondary
eclipse is observed, the possibility of a primary transit will present
itself 48.48 days later which should be sufficient time to schedule
the necessary follow-up resources.


\section{Discussion}

There are various factors which we have not included in this analysis
which we briefly discuss here. Firstly, the effects of transit timing
variations for cases of multi-planet systems has not been considered
in the transit/eclipse predictions. This has been discussed in detail
by several others, such as \citet{ago05} and \citet{hol05}. Attempts
have been made to detect this effect, such as the monitoring of
HD~209458b by the MOST satellite \citep{mil08}, but the effect has not
been observed at the time of writing. For a given transit, the
influence of additional planets on timing variations could be large
depending upon the mass ratio of the planets and the eccentricities of
the orbits. In general though, this effect is relatively small
(minutes) and is cyclic in nature such that the net effect over many
orbits is zero. In addition, many of the more interesting cases
discussed in this paper are highly eccentric giant planets which tend
to be the only known (detectable) planet in those systems.

The second item of note is the issue of how detectable the signature
of a secondary eclipse is, both from the ground and from space. The
attempt to observe a secondary eclipse is undoubtedly much more
difficult than that for a primary eclipse due to such factors as the
relatively low eclipse depth and the wavelength restriction for
optimal detection. Planetary emissions and their associated flux
ratios have been discussed in detail by such papers as \citet{cha05}
and \citet{bur06}. In particular, \citet{bur06} have shown that the
contrast ratios for hot Jupiters in the mid-IR will be of order $>
10^{-3}$. The contrast ratio for longer period planets will scale with
$1/r^2$ where in this case $r$ (as defined in Equation
\ref{separation}) is evaluated at the point of predicted secondary
eclipse. The results described in this paper assist in evaluating the
preferred targets for follow-up and whether a difficult observation
with high probability (secondary eclipse) may be preferable to a
somewhat easier observations with low probability (primary transit),
as was the case for HD~80606b.

Finally, we note that the results of this paper will be verifiable
statistically as more secondary eclipses are detected. This will be
particularly pertinent in upcoming years as the high-precision
photometry from such missions as CoRoT and Kepler are released. One
such secondary eclipse detection has been observed by CoRoT for the
transiting exoplanet CoRoT-2b \citep{alo09}. Though the presence of
the eclipse may be expected for such a short-period planet, the
sensitivity of Kepler to longer period planets will result in the
detection of secondary eclipses without the necessity of primary
transits. The photometric precision of the Kepler photometry is such
that, despite the contrast ratios mentioned above, we can expect many
such detections from the Kepler mission. The equations described in
this paper can then be used to assess if and when the primary transit
will occur.


\section{Conclusions}

Using the known expressions for the a-priori probabilities of primary
and secondary transits, we have confirmed the literature result that
higher eccentricities favor transit and eclipse detections when
compared to circular orbits with the same period. Furthermore, we find
that higher eccentricities produce a larger range of $\omega$ for
which both transit and eclipse probabilities are increased with
respect to corresponding circular orbits. We show the probabilities of
secondary eclipses for the known exoplanets in combination with the
range of periastron arguments for which both eclipse and transit
probabilities are enhanced. Applying our insight to the planets
catalogued by \citet{but06}, we find there are several interesting
cases that warrant further investigation.

Furthermore, we show that the constraints placed upon the orbital
inclination through detection of a secondary eclipse can substantially
improve a-posteriori estimates of transit probability even with weak
constraints upon the impact parameter. For 75\% of the planets
considered here, the transit probability reaches 100\% if a secondary
eclipse is observed. We also provide analytical expressions for
calculating the time from detected secondary eclipse to the time of
predicted primary transit. We present several case studies with
relatively high secondary eclipse probabilities with respect to their
periods. The planets HD~4113b and HD~37605b present analogous
secondary eclipse potential to HD~80606b and we encourage follow-up of
these targets at predicted secondary eclipse times provided the
eclipse windows can be sufficiently constrained.

With the picture of planetary atmospheres becoming gradually clearer,
the importance of additional eclipsing planets is obvious. For planets
with high secondary eclipse probabilities, the high-risk/high-return
strategy of monitoring these planets will reap significant rewards in
the growing field of exoplanetary atmospheres.


\section*{Acknowledgements}

The authors would like to thank Scott Fleming and Suvrath Mahadevan
for several useful suggestions and the anonymous referee for a thorough and expeditious critique of the paper.



\begin{thebibliography}{}

\bibitem[\protect\citeauthoryear{Agol et al.}{2005}]{ago05} Agol, E.,
  Steffen, J., Sari, R., Clarkson, W., 2005, MNRAS, 359, 567
\bibitem[\protect\citeauthoryear{Alonso et al.}{2009}]{alo09} Alonso,
  R., Guillot, T., Mazeh, T., Aigrain, S., Alapini, A., Barge, P.,
  Hatzes, A., Pont, F., 2009, A\&A, in press (arXiv:0906.2814)
\bibitem[\protect\citeauthoryear{Barbieri et al.}{2007}]{bar07a}
  Barbieri, M., et al., 2007, A\&A, 476, L13
\bibitem[\protect\citeauthoryear{Barnes}{2007}]{bar07b} Barnes, J.W.,
  2007, PASP, 119, 986
\bibitem[\protect\citeauthoryear{Burke}{2008}]{bur08} Burke, C.J.,
  2008, ApJ, 679, 1566
\bibitem[\protect\citeauthoryear{Burrows et al.}{2005}]{bur05}
  Burrows, A., Hubeny, I., Sudarsky, D., 2005, ApJ, 625, L135
\bibitem[\protect\citeauthoryear{Burrows et al.}{2006}]{bur06}
  Burrows, A., Sudarsky, D., Hubeny, I., 2006, ApJ, 650, 1140
\bibitem[\protect\citeauthoryear{Butler et al.}{2006}]{but06} Butler,
  R.P., et al., 2006, ApJ, 646, 505
\bibitem[\protect\citeauthoryear{Charbonneau et al.}{2005}]{cha05}
  Charbonneau, D., et al., 2005, ApJ, 626, 523
\bibitem[\protect\citeauthoryear{Cochran et al.}{2004}]{coc04}
  Cochran, W., et al., 2004, ApJ, 611, L133
\bibitem[\protect\citeauthoryear{de Mooij \& Snellen}{2009}]{dem09} de
  Mooij, E.J.W., Snellen, I.A.G., 2009, A\&A, 493, L35
\bibitem[\protect\citeauthoryear{Deming et al.}{2006}]{dem06} Deming,
  D., Harrington, J., Seager, S., Richardson, L.J., 2006, ApJ, 644,
  560
\bibitem[\protect\citeauthoryear{Deming et al.}{2007a}]{dem07a} Deming,
  D., Richardson, L.J., Harrington, J., 2007, MNRAS, 378, 148
\bibitem[\protect\citeauthoryear{Deming et al.}{2007b}]{dem07b}
  Deming, D., Harrington, J., Laughlin, G., Seager, S., Navarro, S.B.,
  Bowman, W.C., Horning, K., 2007, ApJ, 667, L199
\bibitem[\protect\citeauthoryear{Fossey et al.}{2009}]{fos09} Fossey,
  S.J., Waldman, I.P., Kipping, D.M., 2009, MNRAS, 396, L16
\bibitem[\protect\citeauthoryear{Garcia-Melendo \&
    McCullough}{2009}]{gar09} Garcia-Melendo, E., McCullough, P.R.,
  2009, ApJ, 698, 558
\bibitem[\protect\citeauthoryear{Gillon et al.}{2009}]{gil09} Gillon,
  M., 2009, A\&A, in press (arXiv:0905.4571)
\bibitem[\protect\citeauthoryear{Grillmair et al.}{2008}]{gri08}
  Grillmair, C.J., et al., 2008, Nature, 456, 767
\bibitem[\protect\citeauthoryear{Holman \& Murray}{2005}]{hol05}
  Holman, M.J., Murray, N.W., 2005, Science, 307, 1288
\bibitem[\protect\citeauthoryear{Kane}{2007}]{kan07} Kane, S.R.,
  2007, MNRAS, 380, 1488
\bibitem[\protect\citeauthoryear{Kane \& von Braun}{2008}]{kan08}
  Kane, S.R., von Braun, K., 2008, ApJ, 689, 492
\bibitem[\protect\citeauthoryear{Knutson et al.}{2007}]{knu07} Knutson,
  H.A., et al., 2007, Nature, 447, 183
\bibitem[\protect\citeauthoryear{Laughlin et al.}{2009}]{lau09}
  Laughlin, G., Deming, D., Langton, J., Kasen, D., Vogt, S., Butler,
  P., Rivera, E., Meschiari, S., 2009, Nature, 457, 562
\bibitem[\protect\citeauthoryear{Miller-Ricci et al.}{2008}]{mil08}
  Miller-Ricci, E., et al., 2008, ApJ, 682, 586
\bibitem[\protect\citeauthoryear{Moutou et al.}{2009}]{mou09} Moutou
  et al., 2009, A\&A, 498, L5
\bibitem[\protect\citeauthoryear{Naef et al.}{2001}]{nae01} Naef, D.,
  et al., 2001, A\&A, 375, L27
\bibitem[\protect\citeauthoryear{Seager \&
    Mall\'en-Ornelas}{2003}]{sea03} Seager, S., Mall\'en-Ornelas, G.,
  2003, ApJ, 585, 1038
\bibitem[\protect\citeauthoryear{Sing \&
    L\'opez-Morales}{2009}]{sin09} Sing, D.K., L\'opez-Morales, M.,
  2009, A\&A, 493, L31
\bibitem[\protect\citeauthoryear{Tamuz et al.}{2008}]{tam08} Tamuz,
  O., et al., 2008, A\&A, 480, L33
\bibitem[\protect\citeauthoryear{Williams et al.}{2006}]{wil06}
  Williams, P.K.G., Charbonneau, D., Cooper, C.S., Showman, A.P.,
  Fortney, J.J., 2006, ApJ, 649, 1020
\bibitem[\protect\citeauthoryear{Winn et al.}{2009}]{win09} Winn,
  J.N., et al., 2009, ApJ, 693, 794

\end{thebibliography}
\end{document}